\input harvmac
\input epsf

\newcount\figno
\figno=0
\def\fig#1#2#3{
\par\begingroup\parindent=0pt\leftskip=1cm\rightskip=1cm\parindent=0pt
\baselineskip=12pt
\global\advance\figno by 1
\midinsert
\epsfxsize=#3
\centerline{\epsfbox{#2}}
\vskip 14pt

{\bf Fig. \the\figno:} #1\par
\endinsert\endgroup\par
}
\def\figlabel#1{\xdef#1{\the\figno}}
\def\encadremath#1{\vbox{\hrule\hbox{\vrule\kern8pt\vbox{\kern8pt
\hbox{$\displaystyle #1$}\kern8pt}
\kern8pt\vrule}\hrule}}

\overfullrule=0pt

\noblackbox
\parskip=1.5mm

\def\Title#1#2{\rightline{#1}\ifx\answ\bigans\nopagenumbers\pageno0
\else\pageno1\vskip.5in\fi \centerline{\titlefont #2}\vskip .3in}

\font\caps=cmcsc10

\noblackbox
\parskip=1.5mm



           \def\CO{{\cal O}}


\def\dj{\hbox{d\kern-0.347em \vrule width 0.3em height 1.252ex depth
-1.21ex \kern 0.051em}}

\def\half{{1\over 2}\,}

\def\pt{\partial}

\def\Dirac{\,\raise.15ex\hbox{/}\mkern-13.5mu D}
\def\dirac{\,\raise.15ex\hbox{/}\kern-.57em \partial}
\def\aslash{\,\raise.15ex\hbox{/}\mkern-13.5mu A}

\def\shalf{{\ifinner {\textstyle {1 \over 2}}\else {1 \over 2} \fi}}
\def\sshalf{{\ifinner {\scriptstyle {1 \over 2}}\else {1 \over 2} \fi}}
\def\sfourth{{\ifinner {\textstyle {1 \over 4}}\else {1 \over 4} \fi}}
\def\sthreehalfs{{\ifinner {\textstyle {3 \over 2}}\else {3 \over 2} \fi}}
\def\sdhalfs{{\ifinner {\textstyle {d \over 2}}\else {d \over 2} \fi}}
\def\sdmtwohalfs{{\ifinner {\textstyle {d-2 \over 2}}\else {d-2 \over 2} \fi}}
\def\sdmasonehalfs{{\ifinner {\textstyle {d+1 \over 2}}\else {d+1 \over 2} \fi}}
\def\sdmasthreehalfs{{\ifinner {\textstyle {d+3 \over 2}}\else {d+3 \over 2} \fi}}
\def\sdmastwohalfs{{\ifinner {\textstyle {d+2 \over 2}}\else {d+2 \over 2} \fi}}


  \lref\usd{
  J.~L.~F.~Barbon and E.~Rabinovici,
  ``AdS Crunches, CFT Falls And Cosmological Complementarity,''
  JHEP {\bf 1104}, 044 (2011)
  [arXiv:1102.3015 [hep-th]].
  }

\lref\rt{
  S.~Ryu and T.~Takayanagi, ``Holographic derivation of entanglement entropy from AdS/CFT,'' Phys.\ Rev.\ Lett.\ {\bf
    96}, 181602 (2006) [arXiv:hep-th/0603001].

  ``Aspects of holographic entanglement entropy,'' JHEP {\bf 0608}, 045 (2006) [arXiv:hep-th/0605073].}

 \lref\eternalmalda{
 J.~M.~Maldacena,
  ``Eternal black holes in anti-de Sitter,''
  JHEP {\bf 0304}, 021 (2003)
  [hep-th/0106112].
  }
  
  \lref\maldahartman{
  T.~Hartman and J.~Maldacena,
  ``Time Evolution of Entanglement Entropy from Black Hole Interiors,''
  JHEP {\bf 1305}, 014 (2013)
  [arXiv:1303.1080 [hep-th]].
  }
  
  \lref\epr{
   J.~Maldacena and L.~Susskind,
  ``Cool horizons for entangled black holes,''
  Fortsch.\ Phys.\  {\bf 61}, 781 (2013)
  [arXiv:1306.0533 [hep-th]].
  }

  \lref\mvr{
   M.~Van Raamsdonk,
  ``Building up spacetime with quantum entanglement,''
  Gen.\ Rel.\ Grav.\  {\bf 42}, 2323 (2010)
  [Int.\ J.\ Mod.\ Phys.\ D {\bf 19}, 2429 (2010)]
  [arXiv:1005.3035 [hep-th]].
  
  M.~Van Raamsdonk,
  ``A patchwork description of dual spacetimes in AdS/CFT,''
  Class.\ Quant.\ Grav.\  {\bf 28}, 065002 (2011).
  }

  \lref\llast{
   L.~Susskind,
  ``The Typical-State Paradox: Diagnosing Horizons with Complexity,''
  arXiv:1507.02287 [hep-th].
  }
  
  \lref\enen{
  L.~Susskind,
  ``Entanglement is not Enough,''
  arXiv:1411.0690 [hep-th].
  }
  
\lref\stanfordsus{D.~Stanford and L.~Susskind,
  ``Complexity and Shock Wave Geometries,''
  Phys.\ Rev.\ D {\bf 90}, no. 12, 126007 (2014)
  [arXiv:1406.2678 [hep-th]].
}

\lref\bridges{
L.~Susskind and Y.~Zhao,
  ``Switchbacks and the Bridge to Nowhere,''
  arXiv:1408.2823 [hep-th].
  }
\lref\suscomple{
L.~Susskind,
  ``Computational Complexity and Black Hole Horizons,''
  arXiv:1403.5695 [hep-th], arXiv:1402.5674 [hep-th].
  }
 
  \lref\robertsss{D.~A.~Roberts, D.~Stanford and L.~Susskind,
  ``Localized shocks,''
  JHEP {\bf 1503}, 051 (2015)
  [arXiv:1409.8180 [hep-th]].}
  
 \lref\tensors{
 B.~Swingle,
  ``Entanglement Renormalization and Holography,''
  Phys.\ Rev.\ D {\bf 86}, 065007 (2012)
  [arXiv:0905.1317 [cond-mat.str-el]].
  
  G.~Evenbly and G.~Vidal, ``Tensor Network States and Geometry,"
  Journal of Statistical Physics 145 (2011) 891Ð918, [arXiv:1106.1082 [quant-ph]].
  
   B.~Swingle,
   ``Constructing holographic spacetimes using entanglement renormalization,''
  arXiv:1209.3304 [hep-th].
  
  X.~L.~Qi,
  ``Exact holographic mapping and emergent space-time geometry,''
  arXiv:1309.6282 [hep-th].
  
  J.~I.~Latorre and G.~Sierra,
  ``Holographic codes,''
  arXiv:1502.06618 [quant-ph].
  
  F.~Pastawski, B.~Yoshida, D.~Harlow and J.~Preskill,
  ``Holographic quantum error-correcting codes: Toy models for the bulk/boundary correspondence,''
  JHEP {\bf 1506}, 149 (2015)
  [arXiv:1503.06237 [hep-th]].
  }

  \lref\horo{
  N.~Engelhardt, T.~Hertog and G.~T.~Horowitz,
  ``Holographic Signatures of Cosmological Singularities,''
  Phys.\ Rev.\ Lett.\  {\bf 113}, 121602 (2014)
  [arXiv:1404.2309 [hep-th]].
  
  N.~Engelhardt, T.~Hertog and G.~T.~Horowitz,
  ``Further Holographic Investigations of Big Bang Singularities,''
  JHEP {\bf 1507} (2015) 044
  [arXiv:1503.08838 [hep-th]].
  }

\lref\bana{
M.~Banados,
  ``Constant curvature black holes,''
  Phys.\ Rev.\ D {\bf 57}, 1068 (1998)
  [gr-qc/9703040].
  
   M.~Banados, A.~Gomberoff and C.~Martinez,
  ``Anti-de Sitter space and black holes,''
  Class.\ Quant.\ Grav.\  {\bf 15}, 3575 (1998)
  [hep-th/9805087].

}

\lref\maldads{
J.~Maldacena and G.~L.~Pimentel,
  ``Entanglement entropy in de Sitter space,''
  JHEP {\bf 1302}, 038 (2013)
  [arXiv:1210.7244 [hep-th]].
	}

 \lref\lasttaka{
  M.~Miyaji, T.~Numasawa, N.~Shiba, T.~Takayanagi and K.~Watanabe,
  ``Gravity Dual of Quantum Information Metric,''
  arXiv:1507.07555 [hep-th].
  }

\lref\insightful{G.~Horowitz, A.~Lawrence and E.~Silverstein,
  ``Insightful D-branes,''
  JHEP {\bf 0907}, 057 (2009)
  [arXiv:0904.3922 [hep-th]].}

\lref\usu{
  J.~L.~F.~Barb\'on and E.~Rabinovici,
  ``Holography of AdS vacuum bubbles,''
  JHEP {\bf 1004}, 123 (2010)
  [arXiv:1003.4966 [hep-th]].}

\lref\falls{
J.~L.~F.~Barbon and E.~Rabinovici,
  ``AdS Crunches, CFT Falls And Cosmological Complementarity,''
  JHEP {\bf 1104}, 044 (2011)
  [arXiv:1102.3015 [hep-th]].
  }
  
\lref\complmaps{
J.~L.~F.~Barbon and E.~Rabinovici,
  ``Conformal Complementarity Maps,''
  JHEP {\bf 1312}, 023 (2013)
  [arXiv:1308.1921 [hep-th]].
  }  
  
\lref\maldab{
  J.~Maldacena,
  ``Vacuum decay into Anti de Sitter space,''
  arXiv:1012.0274 [hep-th].
}

 \lref\adscft{
  J.~M.~Maldacena,
  ``The large N limit of superconformal field theories and supergravity,''
  Adv.\ Theor.\ Math.\ Phys.\  {\bf 2}, 231 (1998)
  [Int.\ J.\ Theor.\ Phys.\  {\bf 38}, 1113 (1999)]
  [arXiv:hep-th/9711200].
 S.~S.~Gubser, I.~R.~Klebanov and A.~M.~Polyakov,
  ``Gauge theory correlators from non-critical string theory,''
  Phys.\ Lett.\  B {\bf 428}, 105 (1998)
  [arXiv:hep-th/9802109].
 E.~Witten,
  ``Anti-de Sitter space and holography,''
  Adv.\ Theor.\ Math.\ Phys.\  {\bf 2}, 253 (1998)
  [arXiv:hep-th/9802150].
  }

  \lref\bkl{
   V.~A.~Belinsky, I.~M.~Khalatnikov and E.~M.~Lifshitz,
  ``Oscillatory approach to a singular point in the relativistic cosmology,''
  Adv.\ Phys.\  {\bf 19}, 525 (1970).
  }

\lref\damour{
T.~Damour, M.~Henneaux and H.~Nicolai,
  ``Cosmological billiards,''
  Class.\ Quant.\ Grav.\  {\bf 20}, R145 (2003)
  [hep-th/0212256].
  }

\lref\mukundmar{
  D.~Marolf, M.~Rangamani and M.~Van Raamsdonk,
  ``Holographic models of de Sitter QFTs,''
  arXiv:1007.3996 [hep-th].}

\lref\her{
     T.~Hertog and G.~T.~Horowitz,
  ``Towards a big crunch dual,''
  JHEP {\bf 0407}, 073 (2004)
  [arXiv:hep-th/0406134].
   T.~Hertog and G.~T.~Horowitz,
  ``Holographic description of AdS cosmologies,''
  JHEP {\bf 0504}, 005 (2005)
  [arXiv:hep-th/0503071].
}

\lref\wbn{
  E.~Witten,
  ``Instability Of The Kaluza-Klein Vacuum,''
  Nucl.\ Phys.\  B {\bf 195}, 481 (1982).}

\lref\harlowsus{
  D.~Harlow and L.~Susskind,
  ``Crunches, Hats, and a Conjecture,''
  arXiv:1012.5302 [hep-th].}
\lref\har{
  D.~Harlow,
  ``Metastability in Anti de Sitter Space,''
  arXiv:1003.5909 [hep-th].}

\lref\action{
A.~R.~Brown, D.~A.~Roberts, L.~Susskind, B.~Swingle and Y.~Zhao,
  ``Complexity Equals Action,''
  arXiv:1509.07876 [hep-th].
  }

\lref\craps{
A.~Bernamonti and B.~Craps,
  ``D-Brane Potentials from Multi-Trace Deformations in AdS/CFT,''
  JHEP {\bf 0908}, 112 (2009)
  [arXiv:0907.0889 [hep-th]].
  }

\lref\moshen{
 M.~Alishahiha,
  ``Holographic Complexity,''
  arXiv:1509.06614 [hep-th].
  }

\lref\nynn{
 S.~R.~Das, J.~Michelson, K.~Narayan and S.~P.~Trivedi,
  ``Time dependent cosmologies and their duals,''
  Phys.\ Rev.\ D {\bf 74} (2006) 026002
  [hep-th/0602107].
  S.~R.~Das, J.~Michelson, K.~Narayan and S.~P.~Trivedi,
  ``Cosmologies with Null Singularities and their Gauge Theory Duals,''
  Phys.\ Rev.\ D {\bf 75}, 026002 (2007)
  [hep-th/0610053].
  A.~Awad, S.~R.~Das, K.~Narayan and S.~P.~Trivedi,
  ``Gauge theory duals of cosmological backgrounds and their energy momentum tensors,''
  Phys.\ Rev.\ D {\bf 77}, 046008 (2008)
  [arXiv:0711.2994 [hep-th]].
  A.~Awad, S.~R.~Das, S.~Nampuri, K.~Narayan and S.~P.~Trivedi,
  ``Gauge Theories with Time Dependent Couplings and their Cosmological Duals,''
  Phys.\ Rev.\ D {\bf 79}, 046004 (2009)
  [arXiv:0807.1517 [hep-th]].
  }

  \lref\nyn{
 S.~R.~Das, J.~Michelson, K.~Narayan and S.~P.~Trivedi,
  ``Time dependent cosmologies and their duals,''
  Phys.\ Rev.\ D {\bf 74} (2006) 026002
  [hep-th/0602107].
   A.~Awad, S.~R.~Das, S.~Nampuri, K.~Narayan and S.~P.~Trivedi,
  ``Gauge Theories with Time Dependent Couplings and their Cosmological Duals,''
  Phys.\ Rev.\ D {\bf 79}, 046004 (2009)
  [arXiv:0807.1517 [hep-th]].
  }

 \lref\dsmass{
  A.~Buchel,
  ``Gauge / gravity correspondence in accelerating universe,''
  Phys.\ Rev.\  D {\bf 65}, 125015 (2002)
  [arXiv:hep-th/0203041].
 
A.~Buchel, P.~Langfelder and J.~Walcher,
  ``On time-dependent backgrounds in supergravity and string theory,''
  Phys.\ Rev.\  D {\bf 67}, 024011 (2003)
  [arXiv:hep-th/0207214].

}

\lref\bon{

  O.~Aharony, M.~Fabinger, G.~T.~Horowitz and E.~Silverstein,
  ``Clean time-dependent string backgrounds from bubble baths,''
  JHEP {\bf 0207}, 007 (2002)
  [arXiv:hep-th/0204158].
  
  V.~Balasubramanian and S.~F.~Ross,
  ``The dual of nothing,''
  Phys.\ Rev.\  D {\bf 66}, 086002 (2002)
  [arXiv:hep-th/0205290].
 
  S.~F.~Ross and G.~Titchener,
  ``Time-dependent spacetimes in AdS/CFT: Bubble and black hole,''
  JHEP {\bf 0502}, 021 (2005)
  [arXiv:hep-th/0411128].
 
  V.~Balasubramanian, K.~Larjo and J.~Simon,
  ``Much ado about nothing,''
  Class.\ Quant.\ Grav.\  {\bf 22}, 4149 (2005)
  [arXiv:hep-th/0502111].
 
  J.~He and M.~Rozali,
  ``On Bubbles of Nothing in AdS/CFT,''
  JHEP {\bf 0709}, 089 (2007)
  [arXiv:hep-th/0703220].
 
}
 
 \lref\noise{
 J.~L.~F.~Barbon and E.~Rabinovici,
  ``Geometry And Quantum Noise,''
  Fortsch.\ Phys.\  {\bf 62}, 626 (2014)
  [arXiv:1404.7085 [hep-th]].
}


\baselineskip=15pt

\line{\hfill IFT UAM/CSIC-2015-101}

\vskip 1cm

\Title{\vbox{\baselineskip 12pt\hbox{}
 }}
{\vbox {\centerline{Holographic Complexity And 
 }
\vskip10pt
\centerline{Spacetime Singularities}
}}

\vskip 0.5cm

\vskip 0.4cm

\centerline{$\quad$ {\caps
Jos\'e L.F. Barb\'on$^\dagger$
 and
Eliezer Rabinovici$^{\star, b}$
}}
\vskip0.5cm

\centerline{{\sl  $^\dagger$ Instituto de F\'{\i}sica Te\'orica IFT UAM/CSIC }}
\centerline{{\sl  C/ Nicol\'as Cabrera 13,
 Campus Universidad Aut\'onoma de Madrid}}
\centerline{{\sl  Madrid 28049, Spain }}
\centerline{{\tt jose.barbon@csic.es}}

\vskip0.1cm

\centerline{{\sl $^\star$
Racah Institute of Physics, The Hebrew University }}
\centerline{{\sl Jerusalem 91904, Israel}}
\centerline{{\tt eliezer@vms.huji.ac.il}}

\vskip0.1cm

\centerline{{\sl $^b$
Laboratoire de Physique Th\'eorique et Hautes Energies,
Universit\'e Pierre et Marie Curie}}
\centerline{{\sl 4 Place Jussieu, 75252 Paris Cedex 05, France}}

\vskip 1cm

\centerline{\bf ABSTRACT}

 \vskip 0.3cm

 \noindent
 We study the evolution of holographic complexity  in various AdS/CFT models containing cosmological crunch singularities. We find that  a notion of complexity measured by extremal bulk volumes tends to decrease as the singularity is approached in CFT time, suggesting that the corresponding quantum states have simpler entanglement structure at the singularity.

\vskip 1cm

\Date{September 2015}

\vfill

\vskip 0.1cm




\baselineskip=15pt

\newsec{Introduction}

\noindent

Holography implies a geometrical encoding of entanglement structures in Quantum Field Theories (QFTs) having holographic duals \refs{\adscft, \rt}. The inverse problem of recovering emergent geometry from the entanglement structure has been the subject of much recent focus \refs\mvr.   One of the new suggested entries associated with this dictionary is a relation between computational complexity of a state and   the volume of bulk `space' in the holographic dual.  A particularly stimulating example is provided by eternal black holes and their EPR=ER interpretation in terms of thermo-field double states (cf. \refs{\eternalmalda, \maldahartman, \epr}). The holographic complexity at long times is dominated by the `growth of the wormhole' leading to  
 \eqn\rate{
 {dC \over dt} \sim T\,S\;,
 }
 where $T$ is the temperature and $S$ the entropy of the eternal black hole. On general grounds, one expects this behavior to continue up to the Heisenberg time of the black hole state, of order $T^{-1} e^S$, when complexity should level off at a value of order $e^S$, \refs\llast.

 Further motivation for some sort of  complexity/volume relation stems from tensor network representations of various CFT states with  holographic interpretation  \refs{\tensors, \maldahartman}. In this set up,  there is a  natural notion of `complexity' associated to  the size of the tensor network.  A continuum version of this idea is embodied by the formula  (cf. 
  \refs{\suscomple,\stanfordsus, \bridges, \robertsss, \enen}) 
\eqn\ansatz{
C(t)\propto {{\rm Vol}(\Sigma_t) \over G\ell}
\;,}
where $\Sigma_t$ is a codimension-one space-like section  of the bulk with extremal volume, $G$ is the effective Newton's constant in AdS and $\ell$ its curvature radius. The  numerical normalization in \ansatz\  is a somewhat  arbitrary choice, unless we find some {\it ab initio} definition of complexity in the continuum CFT (see \refs\lasttaka, \refs\moshen\  and \refs\action\ for recent discussions in this direction). 

The extremal codimension-one surfaces leading to the result \rate\ probe the interior of the eternal black hole geometry. However, they turn out to be {\it maximal} surfaces,  staying far from the black hole singularity as 
 $t\rightarrow \infty$, because space-like volumes actually get `crunched' at the singularity. This  fact prevents a straightforward association between the large complexity of late-time eternal black holes and the occurrence of a singularity in the bulk. 
 
 In this paper we examine the behavior of \ansatz\ in situations which can be interpreted as representing holographic versions of cosmological singularities,  and yet one can make a good guess at the corresponding state on the dual QFT. One interesting aspect of the models studied is the occurrence of crunch singularities which are `visible' in terms of UV degrees of freedom on the QFT side. A general lesson of our analysis is the tendency of holographic complexity to decrease on the approach to the singularities, as a result of an effective loss of degrees of freedom. The details depend on the time-dependent  balance between infrared and ultraviolet degrees of freedom near the singularity, and we study various cases where this balance is controlled by explicit mass scales or effective spatial dimensions. 

We begin in section 2 with a review of some AdS/CFT constructions of cosmological crunches, and proceed in section 3 to discuss their salient properties from the point of view of a volume/complexity relation.

\newsec{Cosmological Frames And Bulk Singularities}

\noindent

There are a number of strategies to engineer bulk cosmological singularities in AdS/CFT models, some of which we briefly review in this section. 
 In general, the models discussed can be described as time-dependent deformations of CFTs. The deformed Hamiltonian is chosen in such a way that it becomes singular in finite time, while we are still able to construct the dual bulk dynamics as explicitly as possible. 
 
  In order to benefit from the UV completeness of CFTs, we should specify the Hamiltonian deformation in terms of marginal or relevant operators. This leads to two major classes of models. In the first case, either a dimensionless coupling or the background metric where the CFT is defined is given a time dependence. In the second case, there is a time-dependent mass scale.

\subsec{Models Based On Singular CFT Frames}

\noindent

 One implementation  of these ideas is to work with a CFT on a  world-volume frame with a singular time dependence, and study how this singularity is realized by the bulk dynamics. In the previous broad classification, this strategy corresponds to adding a time-dependent marginal operator to the CFT. Among the infinite set of models of this type, one looks for those whose dual bulk dynamics is relatively easy to construct.

An example of this sort is provided by  the Kasner metrics 
\eqn\ks{
ds^2_{\rm CFT} = -dt^2 + \sum_{i=1}^{d-1}  t^{2p_i} \,d x_i^{\,2} \;,}
with $d>3$ and parameters $p_i$ satisfying $\sum p_i =\sum_i p_i^2 = 1$.  These metrics have a curvature singularity at $t=0$ provided at least one of the coefficients $p_i$ is different from $0$ or $1$. The singularity looks like a {\it big crunch} in directions with $p_i >0$ and a {\it big rip} in the directions with $p_i <0$ (of which there is at least one for any $d>3$). The $\sum_i p_i =1$ relation ensures that the total spatial volume vanishes linearly with $t$ at the $t=0$ singularity.  

Despite its anisotropic character, the Kasner frame has the  technical advantage of being Ricci flat, which allows us to write one $(d+1)$-dimensional bulk solution with no extra work, namely  
\eqn\kb{
ds^2 = {dr^2 \over r^2} + r^2 \left( -dt^2 + \sum_{i=1}^{d-1}  t^{2p_i} \,d x_i^{\,2}\right)\;.} 
This solution provides a large-$N$ definition of a certain CFT state on the Kasner frame, which we refer to as the `Kasner state'.
Its global structure resembles an AdS Poincar\'e patch, cut by the singularity $t=0$ on the time-reflection spatial surface,  with additional null singularities at $t=\pm \infty$ (see Fig 1). The absence of a clear `turn-off' of the time-dependence makes the Kasner state somewhat difficult to interpret from the point of view of the CFT. In any case, the simplicity of the bulk metric makes this model rather interesting for explicit calculations (cf. \refs\horo\ for a recent study   this model and \refs\nyn\ and references therein for further work along these lines).

\bigskip
\centerline{\epsfxsize=0.3\hsize\epsfbox{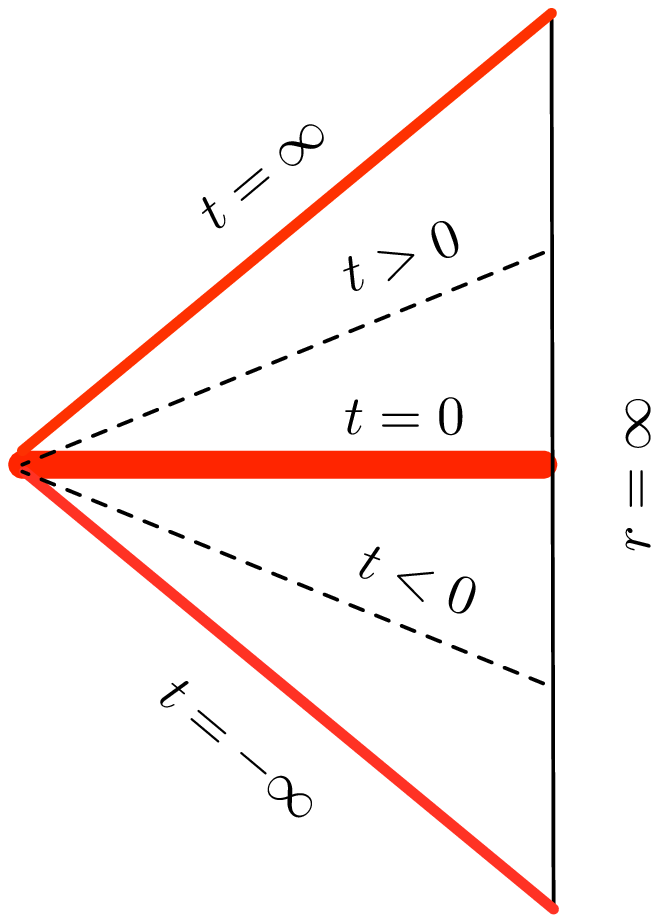}}
\noindent{\ninepoint\sl \baselineskip=2pt {\bf Figure 1:} {\ninerm
Global structure of the bulk dual of the Kasner state, showing a two-dimensional section with $x_i ={\rm constant}$.  }}
\bigskip

Another construction in the same spirit is defined by a CFT on a space-time looking locally like  ${\rm E}_d \times {\bf S}_{R(t)}^1$, where ${\rm E}_d = {\bf R} \times {\bf S}^{d-1}$ is the standard Einstein static universe and the extra ${\bf S}^1$   shrinks to zero radius in a finite time.  The time dependence of the radius,  $R(t)$, is  chosen so that the bulk solution admits a simple parametrization. In particular, working in units of the $(d-1)$-sphere radius, the choice 
\eqn\scc{
ds^2_{\rm CFT} = -dt^2 + d\Omega_{d-1}^2 +  \cos^2 (t) \,d\phi^2\;,
}
with the periodicity  $\phi \equiv \phi + 2\pi R $, corresponds to $R(t) = R \cos (t)$, and  it arises as a conformal  boundary of a periodically identified AdS$_{d+2}$ manifold.  To see this, it is convenient to map the CFT frame by a conformal transformation to ${\rm dS}_d \times {\bf S}^1_R$, where we transfer the time dependence from the circle to a rescaling of the E$_d$ factor, leading to a de Sitter cosmological frame times a fixed circle (cf. Fig. 2)
$$
ds^2_{\rm CFT} = \cos^2 (t)\left[-d\tau^2 + \cosh^2(\tau) \,d\Omega_{d-1}^2 + d\phi^2 \right]\;.
$$
Here we require $dt= d\tau /\cosh(\tau)$, which fixes  the  diffeomorphism $\cosh(\tau) = 1/\cos(t)$ between the two time coordinates $t$ and $\tau$. The metric in square brackets, ${\rm dS}_d \times {\bf S}^1$,
is a conformal boundary metric of
\eqn\bscc{
ds^2_{d+2} = \cosh^2 (\rho) \,d\phi^2 + d\rho^2 + \sinh^2 (\rho) \left(-d\tau^2 + \cosh^2 (\tau)\,d\Omega_{d-1}^2 \right)
\;,}
which is locally the AdS$_{d+2}$ vacuum, up to the identification of the $\phi$ coordinate.\foot{To see this, notice that Wick rotation in the $\tau$ time variable gives a standard representation of Euclidean AdS$_{d+2}$.} 

\bigskip
\centerline{\epsfxsize=0.6\hsize\epsfbox{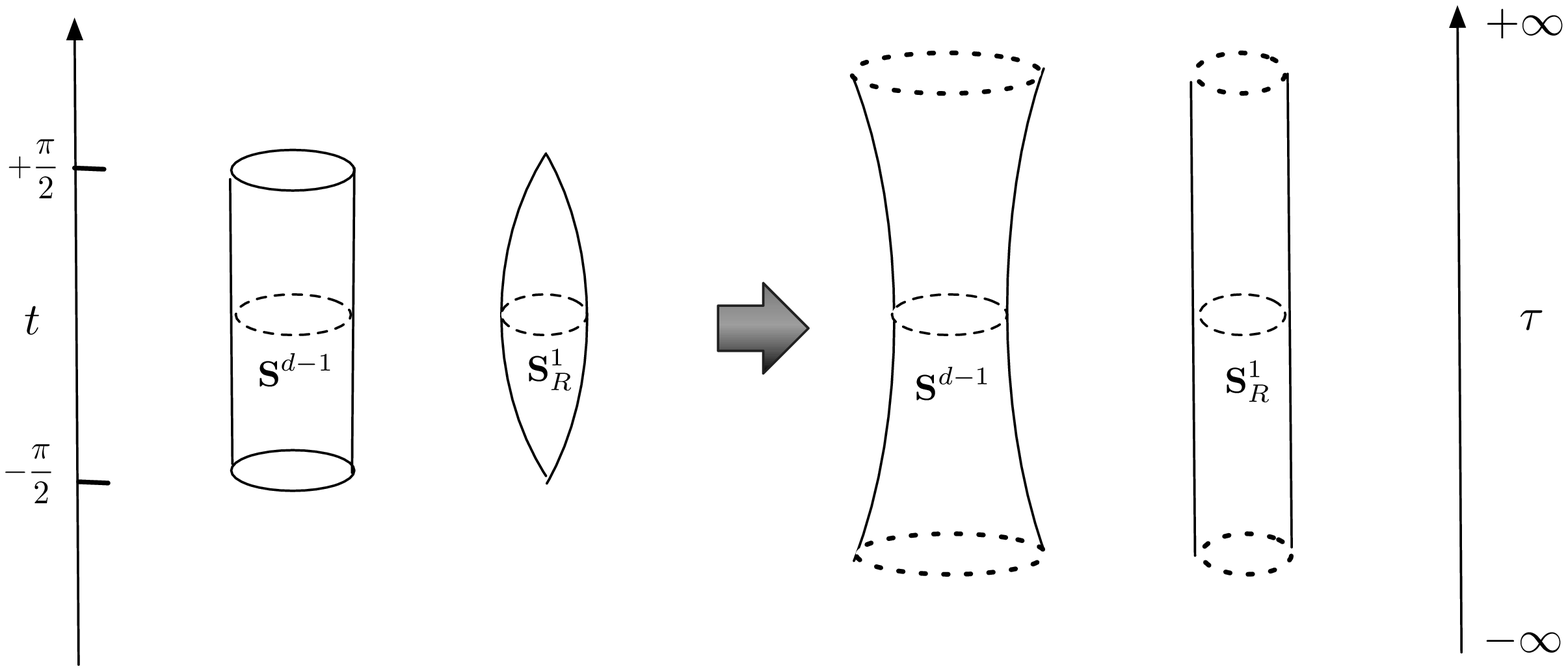}}
\noindent{\ninepoint\sl \baselineskip=2pt {\bf Figure 2:} {\ninerm
The conformal map between the CFT singular frame, ${\bf S}^{d-1}$ times a shrinking circle, and  the `eternal frame' ${\rm dS}_d \times {\bf S}^1_R$. Shown are the projections of time developments along the $(d-1)$-sphere and the circle in both cases.
  }}
\bigskip

The compact nature of $\phi$ produces a global crunch
in the bulk, as seen by continuing \bscc\ past the horizon at $\rho=0$. We can do this by the standard continuation $\rho =i\eta$ and $\tau = \chi -i\pi/2$ which produces a FRW form of the metric:
\eqn\frwc{
ds^2 = -d\eta^2 + \sin^2 (\eta)\left(d\chi^2 + \sinh^2(\chi) d\Omega_{d-1}^2 \right) + \cos^2 (\eta)\,d\phi^2\;,
}
 a circle of radius $R\cos(\eta)$ fibered over the FRW patch of vacuum AdS$_{d+1}$. A singularity arises at $\eta=\pi/2$ where
the circle shrinks to zero size. Since the bulk metric is locally AdS with identifications, we shall refer to this model as the `topological crunch'. A picture of the global structure along the AdS directions is shown in Fig. 3. This model was studied in \refs\bana\ and recently analyzed from the point of view of holographic entanglement entropy in \refs\maldads.\foot{There is a competing solution constructed from the Euclidean AdS$_{d+2}$ black hole metric,  with the same asymptotic ${\rm dS}_d \times {\bf S}^1$ form, where the ${\bf S}^1$ factor is filled in the bulk. This solution has smaller Euclidean action at large $M$, leading to a `bubble of nothing'  \refs{\wbn,\bon} in the bulk and no crunch singularity.}

\bigskip
\centerline{\epsfxsize=0.3\hsize\epsfbox{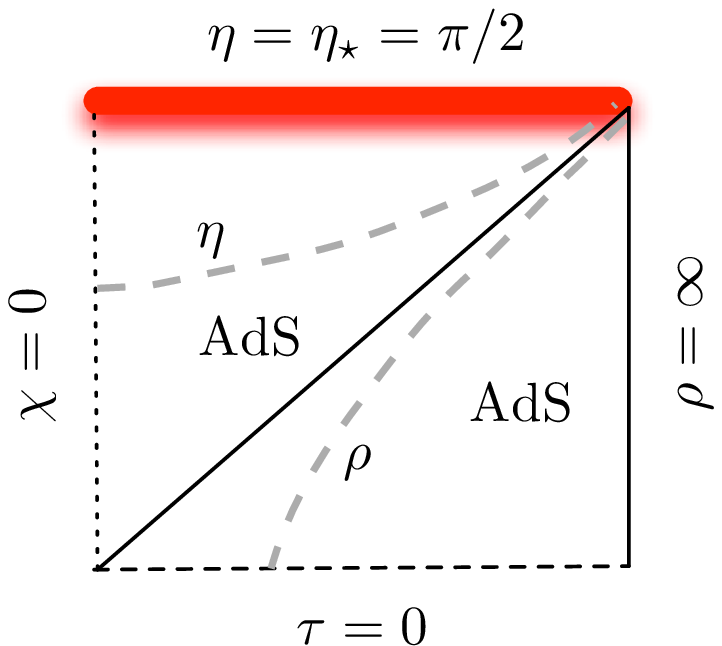}}
\noindent{\ninepoint\sl \baselineskip=2pt {\bf Figure 3:} {\ninerm
Global extension of the AdS section of the topological crunch geometry, showing the constant-$\eta$ space-like surfaces and the constant-$\rho$ time-like surfaces, whose $\rho=0$ limit defines a horizon. Only the future of the time-reflection surface $\tau=0$ is shown. }}
\bigskip

\subsec{de Sitter/Crunch Models}

\noindent

 The topological crunch model was introduced in the previous subsection as a CFT state on the product of the Einstein manifold times a shrinking circle. Alternatively, performing a conformal map we could represent it as a de Sitter-invariant state of the same CFT on dS$_d \times {\bf S}^1$, where now the circle radius, $R$, is constant in de Sitter time. If we look at this model on length scales much larger than $R$, we may integrate out the Kaluza--Klein modes on the ${\bf S}^1$ factor and work with an effective $d$-dimensional effective theory on dS$_d$. This theory has a mass scale $M=(2\pi R)^{-1}$ leading to a tower of irrelevant operators. This suggests that there should exist versions of this set up involving massive deformations of CFTs on $d$-dimensional de Sitter space. If we introduce the mass scale $M$ via a relevant operator, the UV of the field theory is still $d$ dimensional. We then consider  a CFT on dS$_d$, perturbed by a relevant operator $\CO$ of weight $\Delta <d$,  
\eqn\relv{
\delta I = \int_{{\rm dS}_d} M^{d-\Delta} \;\CO\;,
}
 where  $M$ is a fixed mass scale, expressed in units of the Hubble parameter (cf. \refs\dsmass). In terms of standard QFT intuition, the effect of this operator is clearest in the limit $M\gg 1$, since in this case the massive deformation decouples from the Hubble expansion. Although generically we may expect a gapped theory at the scale $M$, we may also  have a non-trivial infrared CFT surviving at distances much larger than $M^{-1}$, which is subsequently placed on an slowly expanding de Sitter space-time. The bulk picture for this scenario is a de Sitter-invariant  background 
 \eqn\sds{
 ds^2_{\rm bulk}  = d\rho^2 + f(\rho)^2 \,ds^2_{{\rm dS}_d}\,
}
  with the warp function $f(\rho)$ depending on $M$. When $M\gg 1$ we expect this function to be well approximated by a narrow domain wall located at a fixed value of $\rho=\rho_M$, separating two almost-vacuum AdS backgrounds $f_\pm (\rho) = \sinh(\rho/\ell_\pm)$ with slightly different curvatures. We take the asymptotic (UV) AdS, denoted 
AdS$_+$, to have radius of curvature $\ell_+ =1$, whereas the infrared one, AdS$_-$, has radius $\ell_- < 1$. Accordingly, standard renormalization-group intuition implies $N^2_+ > N^2_-$ for the effective species numbers, $N^2_\pm = \ell_\pm^{\,d-1} /G$.  In a more microscopic specification of these models, narrow walls between AdS backgrounds can be realized in terms of probe D-branes, along the lines of \refs{\insightful, \craps, \usu}. In these constructions, one has $N_+ - N_- \ll N_+$, so that $\ell_-$ differs from $\ell_+ =1$ by $1/N$ effects. 

The asymptotic AdS$_+$ patch may be reparametrized as 
$$
ds^2_{{\rm AdS}_+} \approx d\rho^2 + \sinh^2(\rho)\left(-d\tau^2 + \cosh^2 (\tau) \,d\Omega_{d-1}^2 \right)
 = -(1+r^2)\,dt^2 + {dr^2 \over 1+r^2} + r^2\,d\Omega_{d-1}^2\;,
$$
using the explicit map 
\eqn\chang{
r = \sinh(\rho) \cosh(\tau)\;, \qquad \cos (t) = {\cosh(\rho) \over \sqrt{1+\sinh^2 (\rho) \cosh^2 (\tau)}}\;.
}
Therefore, the domain wall at $\rho=\rho_M$ will execute a de Sitter-invariant motion of the form 
\eqn\te{
r(t)_{\rm wall} = \left({1+r_M^2 \over \cos^2 (t)} -1\right)^{1/2}\;,
}
in the global coordinate system, where $r_M = \sinh(\rho_M)$ is  interpreted as  the minimum value of $r(t)$. Since the global coordinate system is adapted to the static frame of the CFT, the background \sds\  can be interpreted as a $t$-dependent state on the same CFT defined on the Einstein universe (E-frame) 
\eqn\eu{
ds^2_{{\rm E}_d} = -dt^2 + d\Omega_{d-1}^2\;.}
For $M\gg 1$ we have the UV/IR relation $\sinh (\rho_M) =r_M \approx M$. 
Moreover, at any time such that $r(t)\gg 1$ we can approximate \te\ by  $r(t) \approx r_M /\cos (t)$. Defining a $t$-dependent version of the UV/IR relation in the Einstein frame,  $M(t) = r(t)$, we learn that this background can be interpreted as a deformation of the E-frame CFT by the same relevant operator, but now with a $t$-dependent mass scale 
\eqn\ms{
M(t) = {M \over \cos(t)}\;.}
We could have anticipated this result by noticing that the dS and E frames of the CFT are conformally related, 
$
ds^2_{{\rm dS}_d} = \cosh^2 (\tau) \,ds^2_{{\rm E}_d}$, 
provided we link the time variables according to 
$
dt  = d\tau / \cosh (\tau)
$. So the $t$-dependent mass scale \ms\ arises as the simple consequence of rewriting the relevant perturbation \relv\ in the Einstein frame:
\eqn\relve{
\delta I = \int_{{\rm E}_d} (M(t))^{d-\Delta} \,\CO\;.
}

In the case that the relevant operator leaves no large-$N$ CFT in the infrared, we can expect an  interior region of stringy curvature or perhaps a `bubble of nothing'  \refs\wbn, which effectively cuts off the bulk at $\rho=\rho_M$ (see \refs\mukundmar\ and references therein). Equivalently, the bubble of nothing grows in global coordinates following \te\ above. 

In the case when the perturbation is softer than Hubble,  $M\ll 1$, the effects of the expansion cannot be easily disentangled from those of the relevant operator and the QFT intuition is less clear. In the holographic picture we can still consider the natural extrapolations of these dS-bubble backgrounds:  either bubbles of nothing with minimal radius much smaller than Hubble, or small bubbles with non-trivial interior.  In the second case, since the bubble wall has an acceleration horizon, we can continue the geometry past it,  using a standard parametrization which preserves the de Sitter isometry group, i.e. a hyperbolic FRW model: 
\eqn\buint{
ds^2_{\rm interior} = -d\eta^2 + a(\eta)^2 \left(d\chi^2 + \sinh^2 (\chi)\,d\Omega_{d-1}^2 \right)\;,
} 
with $a(\eta)$ a scale factor satisfying $a(\eta) \approx \eta$ near $\eta=0$ in order to match smoothly at the acceleration horizon of the bubble. In general, the complete solution will require specifying the matter fields $\Phi$ which are dual to the relevant CFT operator $\CO$, and  contribute an extra energy density on top of the AdS vacuum,  implying that $a(\eta)$ may differ significantly from the vacuum form $a(\eta)_{\rm AdS} = \sin(\eta)$.

\bigskip
\centerline{\epsfxsize=0.7\hsize\epsfbox{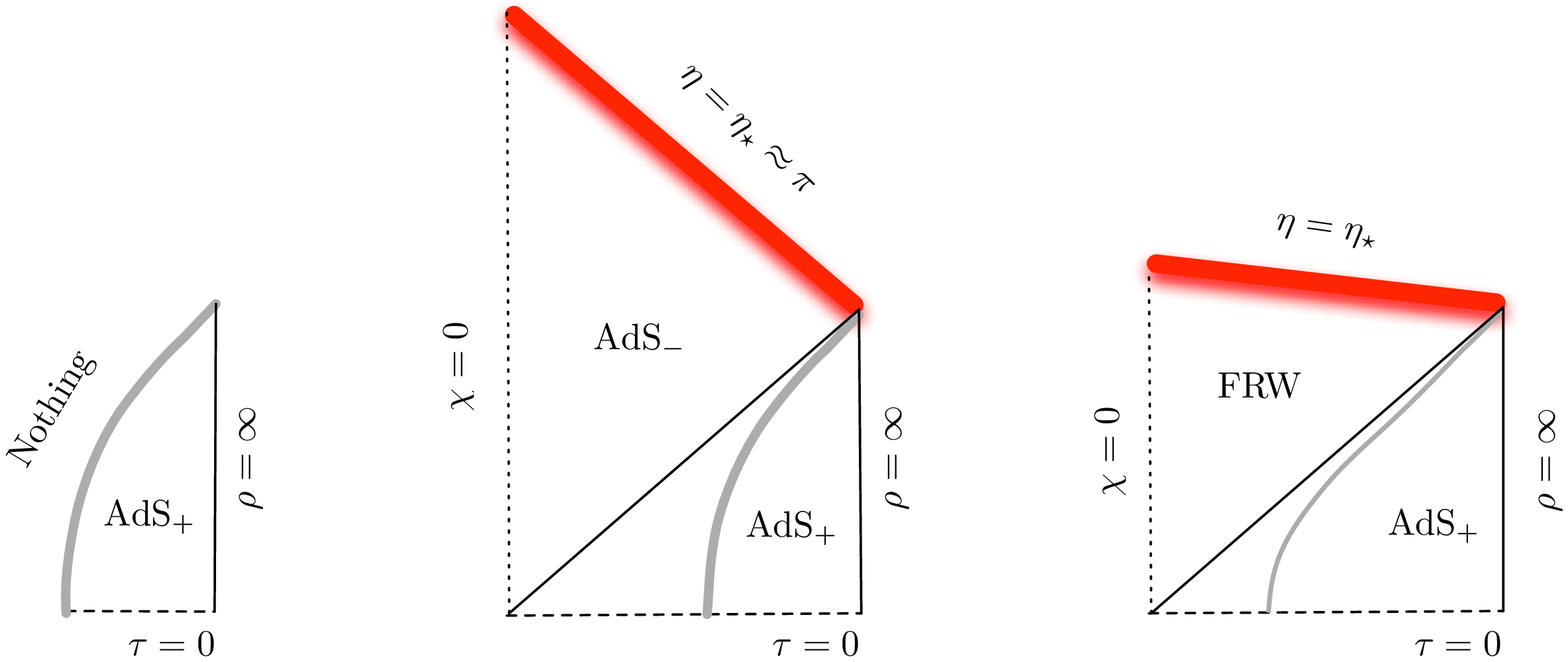}}
\noindent{\ninepoint\sl \baselineskip=2pt {\bf Figure 4:} {\ninerm
Global structure of the (future half) bulk space-time  dual to massive deformations of dS CFTs. All three space-times have vacuum AdS$_+$ on the exterior of a bubble whose wall sits at constant $\rho=\rho_M$ (grey line). On the left we have a dS-invariant bubble with nothing inside, dual to a gapped phase. In the center we have a dS-invariant bubble with an approximate AdS-FRW interior and  thin walls.  On the right we have a smaller dS-invariant bubble with `thick' walls and a more generic negatively curved FRW interior. All these solutions have $O(1,d)$ isometries, and can be obtained as the Lorentzian continuation of Euclidean solutions with AdS asymptotics and $O(d+1)$ symmetry.  }}
\bigskip

Some generic properties of the solution follow from general considerations.  In the simplest situation we can consider a free scalar $\Phi$ of mass $m$, which must be slightly tachyonic  on account of $\CO$ being a relevant operator, i.e. it must satisfy $-d^2 / 4 < m^2 <0$ to preserve the BF bound. Regularity of the scalar field solution at $\rho=\eta=0$ requires $\partial_\eta \Phi |_{\eta =0} =0$, implying that the  energy density contributed by $\Phi$ on FRW hyperbolic slices 
$$
\rho_\Phi = \half (\pt_\eta \Phi)^2 + \half m^2 \Phi^2
$$
starts out {\it negative} at $\eta =0$. Since the FRW model is initially expanding, cosmological friction damps the kinetic energy of the scalar field as it rolls further down the inverted potential, implying that $
\rho_\Phi$ stays negative for some time beyond the turnaround time $\eta_m$. The FRW equation can be written as
\eqn\few{
\left({da \over d\eta}\right)^2 + a^2 -{16\pi G \over d(d-1)} \rho_\Phi \,a^2 -1=0\;,
}
where we have used the normalization $\ell=1$ for the pure AdS solution $a(\eta)_{\rm AdS} = \sin(\eta)$. The negative value of $\rho_\Phi$ translates into a scale factor which stays below the vacuum one, $a(\eta) < \sin(\eta)$, with an earlier turnaround $\eta_m < \pi/2$, and an earlier zero of the scale factor, $a(\eta_\star)=0$, with $\eta_\star < \pi$, which now represents a  curvature singularity. In particular, the maximal scale factor of the hyperbolic FRW sections satisfies $a_m <1$. 
Even the interior geometries with thin walls, corresponding to the $M\gg 1$ case, are expected to induce small tail modifications of the scalar field, which eventually cause the interior FRW solution to differ slightly form AdS$_-$ and ultimately crunch at $\eta_\star \approx \pi$. 

We have collected the different qualitative features of the dS deformed models in Figure 4. A general lesson from these considerations is that the interior of the bubble in dS/Crunch models tends to have {\it smaller} volume than the analogous region in the pure AdS vacuum.  In either case, independently of the ratio of $M$  to the Hubble parameter,  the crunch singularity acquires the QFT interpretation of a  singularity of the driving by the E-frame  relevant operator, characterized by a $t$-dependent mass scale $M(t)$ having a pole at $t=t_\star$. Conversely, the dS-frame description of the same state, which reparametrizes the finite $t$ interval used in the E-frame into an eternal  dS space-time,  gives a completely non-singular description of the physics.  It is in this sense that we can regard the dS-frame description as a `holographic definition' of the crunch singularity (cf. \refs{\maldab, \harlowsus, \falls, \complmaps}).

\newsec{Complexity Estimates}

\noindent

We now turn to evaluate the complexity associated to each of these states, using the suggested formula \ansatz. We parametrize the extremal surfaces $\Sigma_t$ as a function of boundary time coordinates  $t$ for which the singularity is `finitely' far away. In this way we attempt to capture the notion of `complexity of the singularity'.   We focus primarily on the {\it bare}, unrenormalized complexities, defined with a $t$-invariant UV cutoff in place. 

\subsec{Complexity Of The Kasner State}

\noindent

We start with the Kasner state,  rewriting the bulk metric  \kb\ in the form
\eqn\kz{
ds^2 = {1\over z^2} \left(-dt^2 + \sum_i t^{2p_i} dx_i^2 + dz^2 \right)
\;,}
by the change $r=1/z$. 
In computing the complexity of this state,  we let the codimension-one surface extend in the $x_i$ directions and be described by a function $t(z)$ as it enters the bulk from the $z=0$ boundary. The volume functional is then given by (for $t(z)>0$) 
\eqn\volf{
{\rm Vol} (\Sigma_t) = V_x \int {dz \over z^{d}} \,t(z) \,\sqrt{1-t'(z)^2}\;,
}
where $t'(z) = dt /dz$,  $t(0)=t$ is the boundary value, and $V_x$ is the comoving volume in the  $x_i$ coordinates. We can render $V_x$ finite by compactification of all $x_i$ directions into a torus, a situation which makes the null singularities at $t=\pm\infty$ rather evident, since the circles corresponding to negative $p_i$ coefficients shrink to zero size at the Poincar\'e horizon $z=\infty$. In fact, in a string-theory embedding of the problem, light winding modes must be taken into account as soon as the size of these circles becomes of $O(1)$ in string units. In our analysis we shall ignore these effects by taking $V_x$ as formally infinite, and interpreting \volf\ as defining a comoving complexity {\it density}.  
In fact,   \volf\ is always dominated by the small-$z$ region, corresponding to UV degrees of freedom according to the standard AdS/CFT rules. This can be seen by a simple estimate on $t=$ constant surfaces, for which one finds 
\eqn\ves{
{\rm Vol} (\Sigma_{t={\rm constant}}) = V_x \,|t|\,{\Lambda^{d-1} \over d-1}\;,
}
where $\Lambda = 1/z_{\Lambda}$ is the UV cutoff of the CFT. 

The UV dominance of the complexity can be argued on general grounds. Extremal surfaces 
satisfy the Euler--Lagrange equation
\eqn\ela{
{d \over dz} \left[{-t(z)t'(z) \over z^d \sqrt{1-t'(z)^2}}\right] = {\sqrt{1-t'(z)^2} \over z^d}\;.
}
Operating and removing the square roots we obtain the related equation
\eqn\redu{
z\, t \,t'' + z\,(1-t'^2)\,t'^2 -d \cdot t\,t'\,(1-t'^2) + z\,(1-t'^2)^2 =0\;,
}
which always admits the solutions $t' (z) = \pm1$ for any value of the asymptotic limit $t(0)=t$, corresponding to zero volume in \volf. On the other hand, 
near the $z=0$ boundary and for arbitrary values of $t$,   equation \redu\ is solved   by $t(z) \approx $ constant surfaces with the
volume estimate \ves, dominated by the small $z$ region. It turns out that any solution becomes asymptotically null as $z\rightarrow \infty$.  
 We can confirm this by writing the {\it ansatz}
$
t(z) = \pm z + \varepsilon(z)
$
and keeping only terms linear in $\varepsilon (z)$ and its derivatives. We find an equation which only determines $\varepsilon'(z)$, i.e.
$$
z \,\varepsilon'' +2(d-1) \varepsilon' =0\;,
$$
which ensures the null asymptotics as $z\rightarrow \infty$ with small corrections of order $z^{3-2d}$ (cf. Fig. 5). This ensures that all solutions have a UV-dominated volume of the form \ves. 
The corresponding complexity scales as
\eqn\complekasner{
C(t) \sim N^2 V_x\, \Lambda^{d-1} \,|t| \sim N^2 V_{\rm CFT} \,\Lambda^{d-1}\;.
}
That is to say, it scales extensively in the physical volume of the CFT metric, $V_{\rm CFT} = |t|\,V_x$,  which is itself vanishing towards the singularity. 

\bigskip
\centerline{\epsfxsize=0.3\hsize\epsfbox{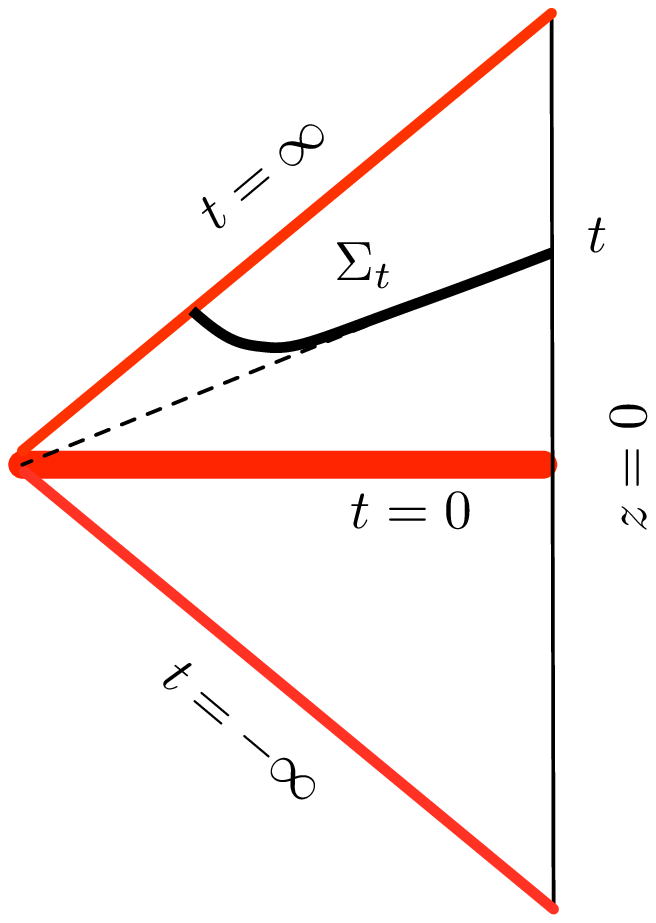}}
\noindent{\ninepoint\sl \baselineskip=2pt {\bf Figure 5:} {\ninerm
Qualitative form of a maximal-volume surface which becomes asymptotically null as $z\rightarrow \infty$ and approaches $t=$ constant at small $z$.  }}
\bigskip

The dominance of the small-$z$ region translates into a complexity proportional to the number of microscopic degrees of freedom, with a time-independent spatial cutoff of order $\Lambda^{-1}$.  It vanishes formally in the $|t|\rightarrow 0$ limit, corresponding to the effective vanishing volume of the Kasner spatial sections.  At any rate, the estimate \complekasner\ should not be applied so close to the singularity that   $V_{\rm CFT} \Lambda^{d-1} <1$, since one is formally left without any `lattice sites' on that comoving volume. Hence, without any more detailed treatment involving for example string dynamics, we can only say that the complexity is extensive in the decreasing  spatial volume, reaching a value of $O(N^2)$ when the region of interest reaches cutoff size.

\subsec{Complexity Of The Topological Crunch}

\noindent

The non-perturbative definition of the topological crunch model is given by a $(d+1)$-dimensional CFT on a fixed $(d-1)$-sphere  times a shrinking circle. Hence, it bears some resemblance with the Kasner state in the sense that there is a singular CFT frame. On the other hand, a full $(d-1)$-dimensional volume survives in this case at the singularity. It is thus  interesting to determine  the corresponding behavior of the holographic complexity. 

Applying the change of variables \chang\  to \bscc, we obtain one  global parametrization of the topological crunch metric with the form 
\eqn\eftb{
ds^2_{d+2} = -(1+r^2) dt^2 + {dr^2 \over 1+r^2} + r^2 \,d\Omega_{d-1}^2 + (1+r^2)\,\cos^2 (t) \,d\phi^2\;.
}
Maximal codimension-one surfaces anchored at fixed boundary time $t$ are not easy to construct analytically, since they wrap a time-dependent circle. There are however, two special cases which admit an exact description. The first  one is the spatial surface $\Sigma_0$ at  the time-symmetric $t=0$ section. It is extremal and has intrinsic AdS$_{d+1}$ geometry, leading to an initial complexity 
\eqn\cin{
C(0) \approx {2\pi R\,\Omega_{d-1} \over G_{d+2}} \int_0^{r_\Lambda} dr \,r^{d-1} 
= {2\pi R\,\Omega_{d-1} \over G_{d+2}} \,{r_\Lambda^d \over d}  \sim N^2\,V_d \, \Lambda^d\;. }
Here, $N^2 \sim 1/G_{d+2}$ is the number of degrees of freedom of the $(d+1)$-dimensional CFT,  $\Omega_{d-1}$ is the volume of a unit ${\bf S}^{d-1}$ sphere, and thus $V_d = 2\pi R\,\Omega_{d-1}$ is the spatial CFT volume at $t=0$. We have identified the CFT cutoff as $r_\Lambda =\Lambda$, using the standard UV/IR relation of AdS/CFT. 

A glance at Fig. 2 shows that the extremal codimension-one surface $\Sigma_\star$, anchored at $t=t_\star = \pi/2$,  lies completely within the causal domain   of the singularity. The late-time surface, being extremal, will be aligned with the isometries of this `interior' region. The change of variables 
\eqn\necha{
\cos(\eta)=\cos(t)\, \sqrt{1+r^2} \;,\qquad \sin(\eta)\,\cosh(\chi) = \sin(t)\, \sqrt{1+r^2} 
}
maps \eftb\ into the interior metric \frwc, whose isometries include translations in the $\chi$ coordinate. Therefore, the late-time extremal surface $\Sigma_\star$ lies along the particular $\eta=\eta_m$ slice which maximizes  the volume factor 
$
\cos(\eta)\,\sin^d (\eta)
$, i.e. $\eta_m$ is fixed by the relation $(d+1) \cos^2 (\eta_m) =1$. The corresponding maximal value of the warp factor is \foot{It is interesting to notice that $f_m <1$ is the analog of the condition $a_m <1$ which was found for dS/Crunch models.}
\eqn\wf{
f_m = \cos(\eta_m) \sin^d (\eta_m) = {d^{d\over 2} \over (d+1)^{d+1 \over 2}} <1\;.
}

\bigskip
\centerline{\epsfxsize=0.4\hsize\epsfbox{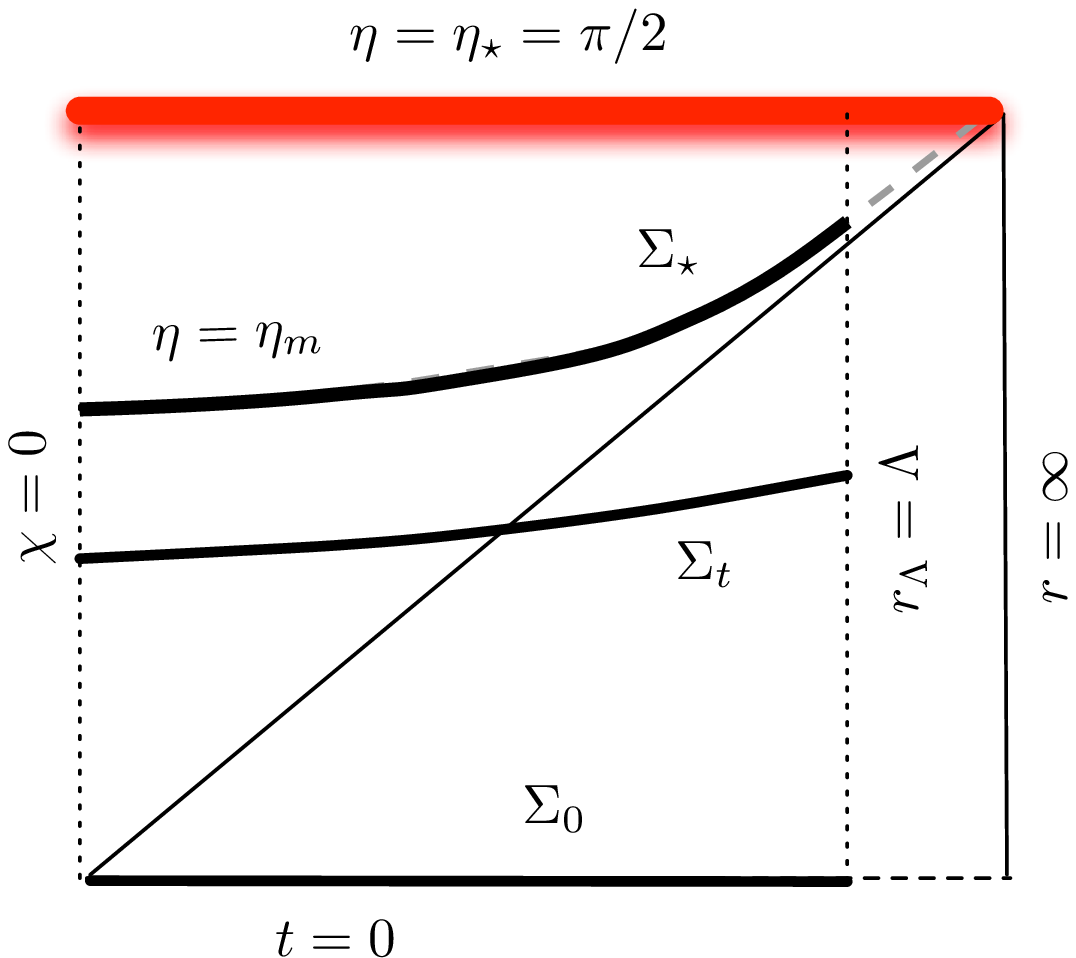}}
\noindent{\ninepoint\sl \baselineskip=2pt {\bf Figure 6:} {\ninerm
Extremal codimension-one surfaces $\Sigma_t$  for the topological crunch model, shown along the AdS section with a UV cutoff $r_\Lambda = \Lambda$ in place. The initial $\Sigma_0$ and final surfaces $\Sigma_\star$ are known to lie along the $t=0$ and $\eta=\eta_m$ surfaces respectively.   }}
\bigskip

In order to compare the initial and final complexities we must regularize the $\Sigma_\star$ surface by cutting it off at the intersection between the $\eta=\eta_m$ and $r=\Lambda$ surfaces (cf. Fig. 6). This determines the limiting value of the $\chi$ coordinate given by the equation
$$
\cosh(\chi_\Lambda) = {\tan (t_\Lambda) \over \tan (\eta_m)}\;,
$$
where $t_\Lambda$ is fixed by the first equation in \necha\ to be
$$
\cos(t_\Lambda) = {\cos(\eta_m) \over \sqrt{1+\Lambda^2}}\;,
$$
resulting in 
$$
\cosh(\chi_\Lambda) \approx {e^{\chi_\Lambda} \over 2} \approx {\Lambda \over \sin(\eta_m)}.
$$

The volume of $\Sigma_\star$ determines the final (regularized) complexity to be 
\eqn\irc{
C(\pi/2) \approx {2\pi R \,\Omega_{d-1} \over G_{d+2}} f_m \int_0^{\chi_\Lambda} d\chi \sinh^{d-1} (\chi) \approx {2\pi R \,\Omega_{d-1} \over G_{d+2}} \, {f_m \over d-1}\, \left({e^{\chi_\Lambda} \over 2}\right)^{d-1} \sim N^2 \,V_d \,\Lambda^{d-1}  \;.
}
We find that the IR complexity is proportional  to the maximal CFT entropy along the non-singular directions.  Despite being macroscopic, this contribution to the complexity is much smaller than the initial one \cin\ because of the smaller power of the cutoff. 
 Hence, we learn that late-time complexity is  smaller than the initial complexity:
\eqn\fcom{
C(\pi/2) < C(0)\;.
}
The topological crunch model behaves in some respects like the Kasner state. Its regularized complexity is dominated at all times by
UV contributions, which are decreasing monotonically towards the singularity as a result of an effective loss of `spatial volume' in the CFT frame. One  new feature is the existence of an infrared contribution to the complexity corresponding to the `interior' portion of $\Sigma_t$ which, while
subleading with respect to the UV contributions, does grow as the singularity is approached and gives the final complexity at $t=t_\star$, a behavior reminiscent of what was found in the case of eternal black holes  \refs{\suscomple,\stanfordsus, \bridges, \robertsss, \enen}.

\subsec{Complexity Of dS/Crunch States}
\noindent

 The case of dS/Crunch models resembles that of the topological crunch in many ways, the main difference being that the bulk is genuinely $(d+1)$-dimensional. Here we distinguish between two qualitatively different situations, corresponding to $M\gg 1$, where the `bubble' has thin walls, and $M\ll 1$, where we have a case of `thick walls'.  

The final extremal surface anchored at the singular time $t=t_\star = \pi/2$ lies in the causal future of the horizon, as seen in Fig. 7. Hence, we have the same phenomenon of a growing infrared component $\Sigma_{\rm IR}$, corresponding to the portion of $\Sigma_t$ lying inside the horizon. It is thus useful to split each surface $\Sigma_t$ into UV and IR components. The asymptotic UV component can be analyzed in terms of an approximate vacuum AdS$_+$ geometry in both cases, whereas the IR component lives in a vacuum AdS$_-$ geometry for $M\gg 1$ and a dynamical FRW geometry for $M\ll 1$. In the first case it is natural to split the UV and IR parts across the `bubble wall' $\rho = \rho_M \gg 1$, as shown in the left panel of Figure 7. In the second case we approximate the late-time surfaces by a $\Sigma_{\rm IR}$ component along $\eta=\eta_m$, plus an asymptotic $\Sigma_{\rm UV}$ surface along a $t={\rm constant}$ slice. The two are matched at the intersection of these space-like surfaces as shown on the right panel of  Figure 7.  

\bigskip
\centerline{\epsfxsize=0.7\hsize\epsfbox{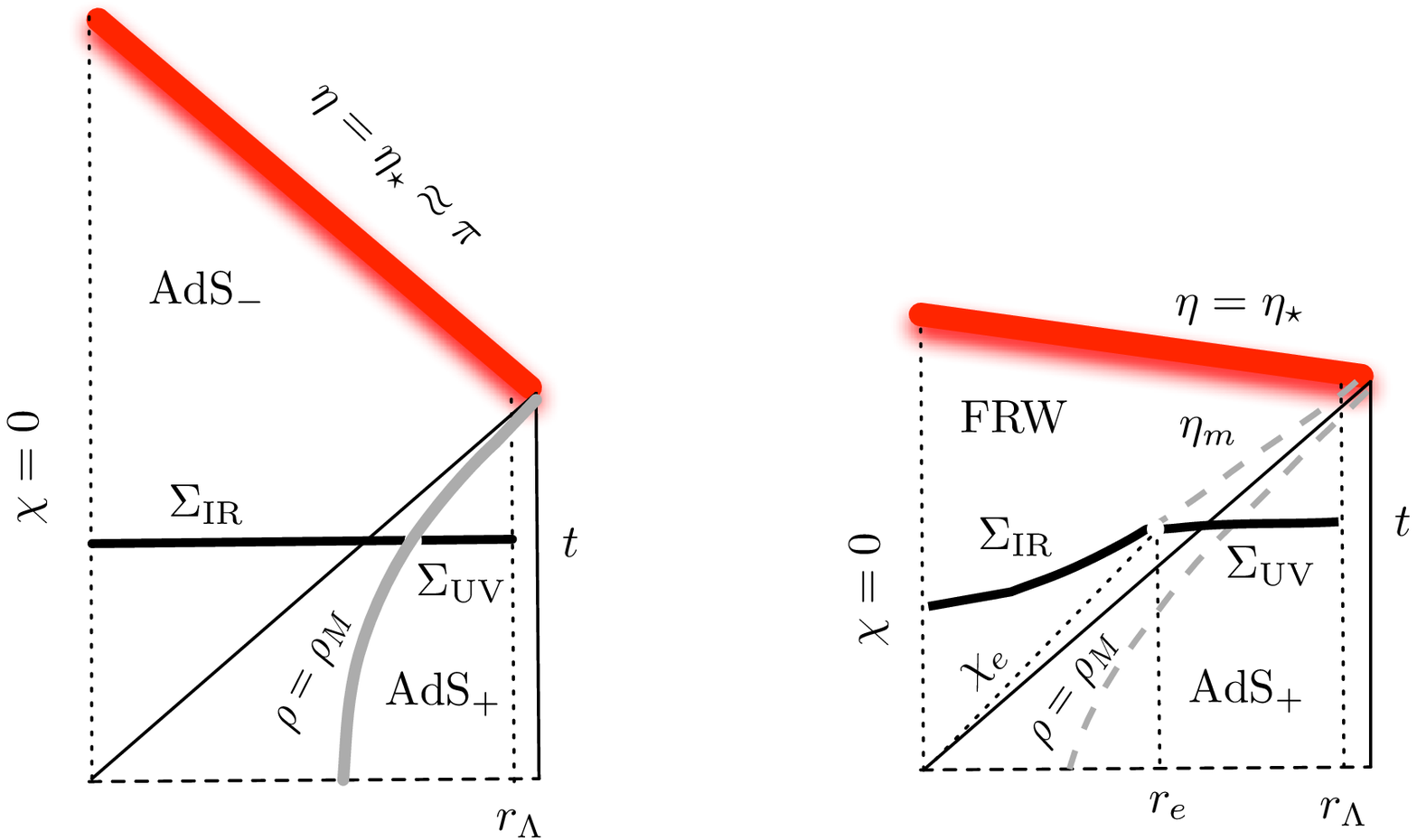}}
\noindent{\ninepoint\sl \baselineskip=2pt {\bf Figure 7:} {\ninerm
Regularized  codimension-one surfaces  for dS/Crunch models with $M\gg 1$ (left) and $M\ll 1$ (right).  The surfaces  are anchored at a particular codimension-one surface of the boundary, labeled by Einstein frame time $t$,  and can be analyzed in terms of  two qualitatively different components $\Sigma_t = \Sigma_{\rm UV} \cup \Sigma_{\rm IR}$.}}
\bigskip

We begin with the analysis of the  $M\gg 1$ case. The UV surface at fixed $t$  extends along the radial interval $r(t)_{\rm wall} < r < r_\Lambda =\Lambda$ with $r (t)_{\rm wall} \approx M(t) = M/\cos(t)$, according to \te.  
Its volume gives a UV contribution to the complexity of order 
\eqn\uvc{
C_{\rm UV} (t) = {\Omega_{d-1} \over G} \int_{r(t)_{\rm wall}}^\Lambda {dr\, r^{d-1} \over \sqrt{1+r^2}}\;,
}
where $\Omega_{d-1}$ is the volume of the ${\bf S}^{d-1}$ sphere, giving the spatial volume $V$ in the CFT metric.  For $M\gg 1$ we can approximate $1+r^2 \approx r^2$ over the whole domain of integration and we have
\eqn\uvcmm{
C_{\rm UV}(t)_{M\gg 1} \approx {\Omega_{d-1} \over G(d-1)} \left(r_\Lambda^{d-1} - r(t)^{d-1}\right) \sim N^2_+ V\,\left(\Lambda^{d-1} - M(t)^{d-1}\right)\;.
}
The result is physically sensible, being  proportional to $S_\Lambda - S_{M(t)}$, where $S_T$ is the high-temperature entropy at temperature $T$. It is always a decreasing function of  E-frame time, since the scale $M(t)$ separating UV from IR degrees of freedom is itself increasing.

If the relevant operator gaps the system at the scale $M$, or leaves out an infrared theory with few degrees of freedom, \uvcmm\ gives the complete answer for the complexity to leading order at large $N_+$.  If an interior AdS$_-$ geometry survives, there is an analogous infrared contribution of the form 
\eqn\irc{
C_{\rm IR} (t) = {\Omega_{d-1} \over G}\int_0^{r(t)_{\rm wall}} {dr\, r^{d-1} \over \sqrt{1+ r^2 / \ell_-^2}}\;,}
where we have included the effect of larger curvature in the interior AdS, i.e. $\ell_- < \ell_+ =1$. Since $r(0) =r_M \approx M \gg 1$, we can approximate the integral by 
$$
C_{\rm IR} (t) \approx {\Omega_{d-1} \ell_- \over G (d-1)} r(t)^{d-1}
$$
The presence of $\ell_-$ in place of $\ell_+ =1$ means that the overall factor of the IR complexity is slightly smaller than the UV one, consistent with the fact that the relevant operator does leave less degrees of freedom in the IR. Hence, we find that the total complexity {\it decreases} as we approach the singularity in the asymptotic time variable. At the final cutoff time, the complexity is given by the IR part,
and scales proportionally to the maximal entropy of the IR CFT.

For $M\ll 1$, the bubble is `small' in AdS$_+$, so we approximate the geometry as AdS$_+$ down to the $\eta=\eta_m$ surface in the interior. Since we argued in section 2 that the interior FRW geometry has `smaller' spatial sections than the vacuum AdS$_+$ geometry,  this approximation provides an upper bound to the UV contribution to the complexity. The split between UV and IR contributions is then determined by the  point of intersection between  the $\eta=\eta_m$ surface and the $t={\rm constant}$ surface, as shown in the right panel of Fig. 7. The value of the  coordinates, $r_e$ and $\chi_e$, is found from the relations \necha,
\eqn\newco{
 \tan(t) = \tan(\eta_m) \,\cosh(\chi_e)\;, \qquad r_e = \sqrt{{\cos^2 (\eta_m) \over \cos^2(t)} -1}\;.
 }
 At large times we can approximate these relations by $e^{\chi_e} /2 \rightarrow \tan(\eta_m)/ \cos(t)$ and $r_e \rightarrow 1 /\tan(\eta_m) \cos(t)$. Hence, we have an upper bound on the late-time UV contribution 
\eqn\uvcmmm{
 C_{\rm UV}(t)_{M\ll 1} \leq {\Omega_{d-1} \over G(d-1)} \left(r_\Lambda^{d-1} - r_e^{d-1}\right) = {\Omega_{d-1} \over G(d-1)} \left(\Lambda^{d-1} - \left({\cos(\eta_m) \over\cos(t)}\right)^{d-1}\right)\;,
}

The infrared component, 
$\Sigma_{\rm IR}$, extends from $\chi=0$ to $\chi=\chi_e$, along the  $\eta=\eta_m$ maximal surface, contributing an `infrared complexity'   at late times of order 
\eqn\irc{
C_{\rm IR} (t)     \approx {a_m^d \Omega_{d-1} \over G} \int_0^{\chi_{e}} d\chi \sinh^{d-1}  (\chi) \approx { \Omega_{d-1}\over G (d-1)} \left({\cos(\eta_m) \over \cos(t)}\right)^{d-1}  \, \left({a_m \over \sin(\eta_m)}\right)^{d-1} \,a_m \;.}
Hence, the IR contribution has the same form as the second term in \uvcmmm, with opposite sign, the precise cancellation only occurring for the  case of pure AdS geometry. When the two contributions are added, the total complexity is a {\it decreasing} function of time, since the
extra factors featuring in \irc\  amount to a numerical coefficient less than unity, as a consequence of the relation $a_m < \sin(\eta_m) <1$, proven in the previous section. 

We conclude that dS/crunch models resemble the topological crunch model in featuring an {\it increasing} infrared complexity, associated to the horizon, and a  {\it decreasing} bare complexity, i.e. $C(0) > C(\pi/2)$. Notice however that these models remain $d$-dimensional  all the way to the singularity and the
complexity is always of order $\alpha(t) \,N^2 V \Lambda^{d-1}$, with $\alpha(t)$ a function decreasing by an $O(1)$ factor between 
$t=0$ and $t_\Lambda= \pi/2 - O(\Lambda^{-1})$.

\subsec{A Comment About Frames}

\noindent

The general lesson of our analysis so far is the tendency of the {\rm bare} holographic complexity to decrease towards singularities.  It is important to notice that this does depend on the frame used to describe the CFT dynamics. In particular, we have emphasized a CFT Hamiltonian picture in which the singularity is met in finite time. If we use an `eternal' frame, we get different answers because of the different weight of UV contributions. Technically, the reason for this is that different frames have different natural cutoffs.  Holographically, we impose the cutoff by declaring $r_f\leq r_\Lambda$ for a radial coordinate $r_f$  such that the metric is asymptotically of the form
$$
ds^2 \Big |_{r_f \rightarrow\infty}  \longrightarrow {dr_f^2 \over r_f^2} + r_f^2 \,ds^2_{\rm CFT\;frame}\;,
$$
and  the particular $r_f$ coordinate does depend on our choice of conformal representative for  the CFT metric. A good example of this phenomenon are the de Sitter models. A fixed de Sitter-invariant cutoff ${\widetilde \Lambda}$ in the dS-frame CFT is implemented holographically as  a fixed-$\rho$ cutoff along a time-like surface   $\rho=\rho_{\tilde \Lambda}= \log({\widetilde \Lambda})$, as shown in Fig. 8. This cutoff surface is    time-dependent with respect to the Einstien-frame time, so that we  would replace $r_\Lambda \sim \Lambda$ in the previous estimates by 
$$
r_{\tilde \Lambda} (t) \sim {{\widetilde \Lambda} \over \cos(t)} \;.
$$
This results in UV contributions to complexity scaling proportionally to ${\widetilde V}_{\rm CFT} {\widetilde \Lambda}^{d-1}$ for the $d$-dimensional models and  ${\widetilde V}_{\rm CFT} {\widetilde \Lambda}^{d}$ for the topological crunch model. Since the dS-frame volume ${\widetilde V}_{\rm dS} $ `inflates' in dS time as a power of $\cosh(\tau)$, the corresponding complexity always increases in this frame. The de Sitter symmetry of the full bulk solutions in both $d$-dimensional models and the topological crunch imply that also the IR contribution scales extensively with ${\widetilde V}_{\rm dS}$. 

\bigskip
\centerline{\epsfxsize=0.3\hsize\epsfbox{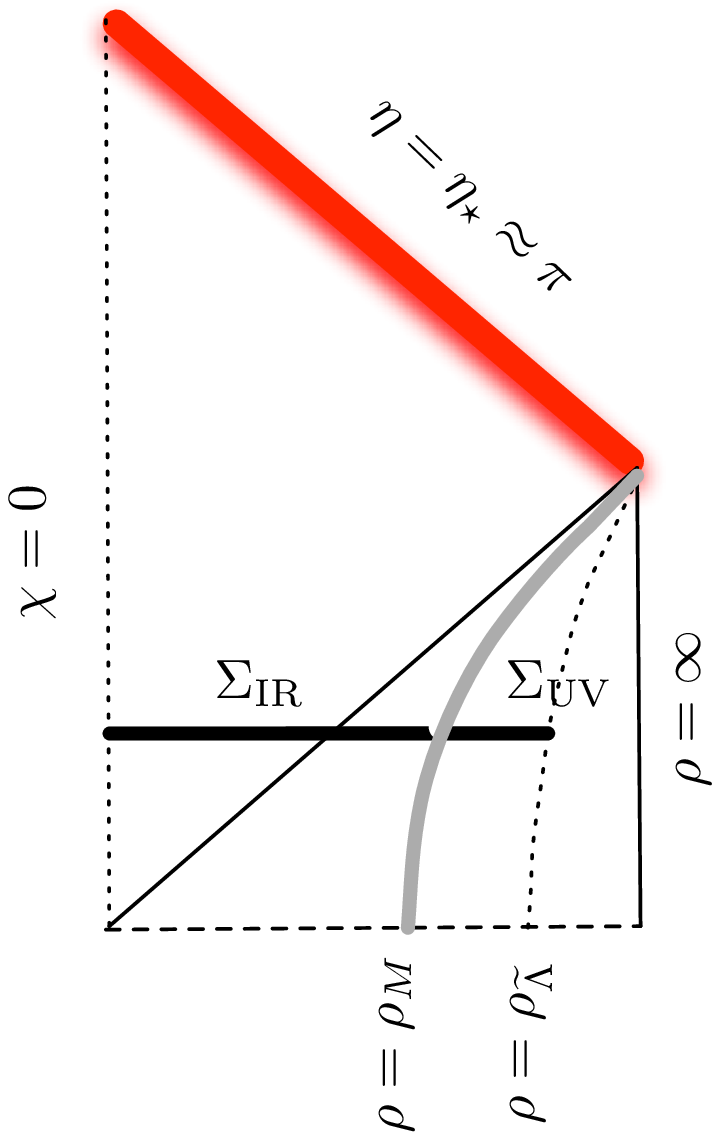}}
\noindent{\ninepoint\sl \baselineskip=2pt {\bf Figure 8:} {\ninerm
A de Sitter-invariant cutoff surface $\rho = \rho_{\tilde \Lambda}$ is $t$-dependent, leading to a {\it growing} volume for the $\Sigma_{\rm UV}$ component resulting from such a regularization. We show here the case of a dS/crunch model with thin walls. }}
\bigskip

The `eternal frame' trick is fairly generic. We can also put the Kasner singularity at the end of an eternity by the change $t=\pm e^{\pm\tau_\pm} $, were the  $\pm$ sign correlates with the bang/crunch phase, namely the crunch sits at the  $\tau_- \rightarrow +\infty$ eternal future, whereas the bang appears at the $\tau_+ \rightarrow -\infty$ eternal past (see \refs\horo). The eternal Kasner frame is related to the standard one by a conformal rescaling  of all lengths by a factor of $1/|t|$. The complexity still scales like  ${\widetilde V}_{\rm CFT} {\widetilde \Lambda}^{d-1}$ in the new frame, but now ${\widetilde V}$  does blow up as we approach $\tau_\pm \rightarrow \mp \infty$.

Since both UV and IR complexities grow without limit in eternal frames, this brings up the question raised in \refs\llast, namely  the existence of non-perturbative  upper bounds on the computational complexity. Those considerations do not apply directly to any of the models in this paper, since entropies also diverge in the eternal frames proportionally to ${\widetilde V}$. Hence, there is no precise sense in which the state wanders inside a finite-dimensional subspace of the Hilbert space,  and no expectations of a complexity bound at long times. The same is true regarding {\it lower} bounds on the long-time behavior of correlation functions (cf. \refs\noise\  and references therein for a recent summary). 

\newsec{Discussion}

\noindent

We have analyzed the behavior of extremal  volumes in a number of AdS cosmologies with a controlled CFT interpretation. Unlike the singularities cloaked inside black hole horizons, these models harbor singularities which are visible to any freely-falling observer in the bulk space-time. In the language of the AdS/CFT correspondence, one can say that these singularities can be efficiently probed by local operators 
in the dual field theory. We have performed estimates in three classes of scenarios and found a tendency for the so-defined holographic complexity to {\it decrease} as the singularity is approached in an appropriate time variable. Taken at face value, this means that the quantum state acquires a simpler entanglement structure when closing into the singularity. A possible interpretation is that tensor network approximations to the wave functions will lose `entangling tensors' as the singularity is approached. It should be stressed that we only establish this monotonic  decrease for times which are sufficiently close to the singularity, but still within the realm of applicability of semiclassical gravity methods. A more accurate analysis would require a microscopic description involving string theory in the bulk and towers of high-dimension operators in the CFT. 

The mechanism for this phenomenon is interesting. Holographically, complexity is identified with a certain maximal-volume space-like slice,  so 
standard crunch singularities tend to `repel' the extremal surface, suggesting that such deep-bulk contributions to holographic  complexity should not decrease near crunch singularities. A time slicing which forces the vanishing volume by getting close to the singularity must probably require an extremely non-local Hamiltonian in the dual QFT picture.  In models were the near-singularity complexity is dominated by the UV, like the Kasner crunch model, the small bulk volume is instead achieved by the  extremal surface becoming approximately null in the IR, and the decreasing complexity can be interpreted as the effect of the volume depletion on the CFT side.  In models where a calculable IR contribution does grow with time, with a behavior similar to that of eternal black holes,  it turns out that the IR/UV interface  is time-dependent, reflecting the fast conversion of UV modes into IR modes. In this process, the IR contribution loses out because of the general properties of the renormalization group, ensuring that some degrees of freedom are always lost in going from the UV to the IR. 

This result contrasts to some extent with the intuition gained by the propagation of probes near crunch singularities, where the high energy blueshift causes unbounded local excitation: singularities are regarded as `hot' and full of complicated dynamics. A generic crunch is usually imagined as a complicated pattern of colliding black holes. On the other hand, the classic BKL work \refs\bkl\ shows that there are universal features to singularities at the ultra-local level (see also the modern work \refs\damour). It is hard to asses the {\it a priori} relevance of this fact, since it depends to a large extent on the particular dynamics of Einstein gravity, but it is tempting to advance a possible relation between the BKL universality and the low holographic complexity found here.

In most of the entries of the AdS/CFT dictionary, at least one side of the duality correspondence is well
defined. In the case at hand both sides are not yet free of ambiguities at this stage. 
On the boundary QFT side, which is supposed to be the responsible adult of this relationship, the definition
of complexity suffers at this stage from various ambiguities, such as the choice of reference state defined conventionally as `simple', or the precise mathematical expression in QFT field variables.  In our treatment the reference state dependence was relegated to the precise UV regulator we
used, namely we only deal with bare complexities, analogous to the size of cutoff tensor networks. Therefore, our implicit choice of reference state is a quantum CFT state which remains disentangled on distance scales larger than the `lattice spacing' $\Lambda^{-1}$, its formal gravity dual corresponding to the $t=0$ section of a  `bubble of nothing' of radius $r_\Lambda = \Lambda$ in AdS. 

 The ambiguities on the
bulk side include the continuing search for criteria to decide on an appropriate bulk quantity and to show
to what it corresponds to. In principle, different bulk quantities should each correspond to some boundary
properties.

\bigskip{\bf Acknowledgements:} 

J.L.F. Barbon wants to thank the LPTHE at Universit\'e Pierre et Marie Curie for hospitality and the Fondation de l'Ecole Normale 
Superieure for partial support, as well as the  MINECO and FEDER  grant  FPA2012-32828, and the 
spanish MINECO {\it Centro de Excelencia Severo Ochoa Program}  grant SEV-2012-0249. 

 E. Rabinovici wants to thank KITP for hospitality during the KITP Workshop 
``Quantum Gravity Foundations: UV to IR" as well as the support of the  {\it Simons Distinguished Professor Chair at KITP}, and the {\it Chaires Internationales de Recherche Blaise Pascal},  financ\'ee par l' Etat et la Region d'Ille-de-France, g\'er\'ee par la Fondation de l'Ecole Normale 
Superieure. His work is also partially supported by the
American-Israeli Bi-National Science Foundation,  the Israel Science Foundation Center
of Excellence and the I Core Program of the Planning and Budgeting Committee and The Israel 
Science Foundation ``The Quantum Universe". 

{\ninerm{
\listrefs
}}

\end